\documentstyle[]{l-aa}

\begin{document}

\thesaurus{02(08.12.2; 10.07.2; 10.08.1; 12.04.1)}
\title{Low-mass stars and star clusters in the dark
Galactic halo}
\author{E.J. Kerins}
\institute{Observatoire de Strasbourg, 11 Rue de l'Universit\'e, 
F-67000 Strasbourg, France.}
\date{Received date / accepted date}
\offprints{{\tt kerins@wirtz.u-strasbg.fr}}
\maketitle

\newcommand{\sm}{\mbox{M}_{\sun}}
\newcommand{\tst}{\textstyle}
\newcommand{\be}{\begin{equation}}
\newcommand{\ee}{\end{equation}}

\begin{abstract}
In a previous study it was proposed that the Galactic dark matter
being detected by gravitational microlensing experiments such as MACHO
may reside in a population of dim halo globular clusters comprising
mostly or entirely low-mass stars just above the hydrogen-burning
limit. It was shown that, for the case of a standard isothermal halo,
the scenario is consistent not only with MACHO observations but also
with cluster dynamical constraints and number-count limits imposed by
20 Hubble Space Telescope (HST) fields. The present work extends the
original study by considering the dependency of the results on halo
model, and by increasing the sample of HST fields to 51 (including the
Hubble Deep Field and Groth Strip fields).  The model dependency of the
results is tested using the same reference power-law halo models
employed by the MACHO team. For the unclustered scenario HST counts
imply a model-dependent halo fraction of at most $0.5-1.1\%$ (95\%
confidence), well below the inferred MACHO fraction. For the cluster
scenario all the halo models permit a range of cluster masses and
radii to satisfy HST, MACHO and dynamical constraints. Whilst the
strong HST limits on the unclustered scenario imply that at least 95\%
of halo stars must reside in clusters at present, this limit is
weakened if the stars which have escaped from clusters retain a degree
of clumpiness in their distribution.
\end{abstract}

\section{Introduction} \label{s1}

The abundance and nature of dark matter in the halo of our Galaxy is
rapidly making the transition from theoretical hypothesis to
observational science. This has been facilitated by the deep
surveys that are now achievable with instruments such as the Hubble
Space Telescope (HST) and by several gravitational microlensing
searches that are currently in progress.

The 2nd-year results from the MACHO microlensing experiment (Alcock et
al. \cite{alc97}) towards the Large Magellanic Cloud (LMC), a
direction which is sensitive to lenses residing in the dark halo,
indicate that a substantial fraction of the halo ($20-100\%$)
comprises objects with a typical mass in the range
$0.1-1~\sm$.\footnote{This conclusion can be avoided if one instead
attributes the observations to lenses residing in a very massive disc,
though to explain the MACHO results one requires a local disc column
density in excess of that typically inferred from kinematical
observations (Kuijken \& Gilmore \cite{kui89}; Bahcall et
al. \cite{bah92}).} These results appear to be broadly supported by
the provisional findings from 4 years of MACHO observations (Axelrod
\cite{axe97}) which have uncovered at least 14 LMC microlensing
candidates. Similar mass scales have also been implicated by the EROS~I
microlensing experiment (Renault et al. \cite{ren97}), though with a
somewhat lower inferred halo fraction. These results are consistent
with the lenses being in the form of low-mass hydrogen-burning stars
or white-dwarf remnants.

However, both of these candidates appear unattractive when other
observational and theoretical results are taken into
consideration. The number density, age and mass function of white
dwarfs in the Galaxy is strongly constrained by number counts of
high-velocity dwarfs in the Solar neighbourhood, and by their helium
production (Carr et al. \cite{carr84}; Ryu et al. \cite{ryu90}; Adams
\& Laughlin \cite{adam96}; Chabrier et al. \cite{chab96}; Graff et
al. \cite{graf97}). In particular, Chabrier et al. (\cite{chab96})
find that a halo fraction compatible with MACHO results requires that
white dwarfs be older than 18 Gyr, though more recently Graff et
al. (\cite{graf97}) have argued for a lower limit closer to 15.5 Gyr
based upon reasonable white-dwarf model assumptions and a halo
fraction of 30\%. The situation for low-mass stars appears at least as
pessimistic with recent HST results indicating that a smoothly
distributed population of low-mass stars can contribute no more than a
few percent to the halo dark matter density, regardless of stellar
metallicity (Bahcall et al. \cite{bah94}; Graff \& Freese
\cite{graf96}; Flynn et al. \cite{fly96}; Kerins \cite{ker97}).

It has been suggested (Kerins \cite{ker97}, hereafter Paper~I) that if
low-mass stars are clumped into globular-cluster configurations then
HST limits can be considerably weakened, since this introduces large
fluctuations in number counts and also may prevent a significant
fraction of sources within the cores of clusters from being
resolved. Motivation for the cluster scenario comes from the
predictions of some baryonic dark matter formation theories, which are
discussed in Paper~I. However, such clusters are required to have
masses and radii consistent with existing dynamical constraints on
clusters and other massive objects residing in the halo. In Paper~I it
was shown that agreement between HST counts, dynamical limits and the
central value for the halo fraction inferred by MACHO (40\% for the
halo model assumed) is possible if clusters have a mass around $4
\times 10^4~\sm$ and radius of a few parsecs. However, HST, MACHO and
dynamical limits are all dependent upon the unknown halo distribution
function, so these results are valid only for the
spherically-symmetric, cored isothermal halo model adopted in Paper~I.

In this paper the model dependency of such conclusions is investigated
using the same set of reference halo models employed in the MACHO
collaboration's analysis of its results. One of these models is
similar, though not identical, to the model investigated in Paper~I,
whilst the other models are constructed from the self-consistent set
of `power-law' halo models presented by Evans (\cite{evan94}). All
models are normalised to be consistent with observational constraints
on the Galactic rotation curve and local column surface density. New
data from the Hubble Deep Field (Flynn et al. \cite{fly96}) and Groth
Strip (Gould et al. \cite{gou97}) are also incorporated, as well as
two other new fields analysed by Gould et al., extending the analysis
from 20 HST fields in Paper~I to 51 in this study.

\section{Halo models}

In Paper~I constraints on the halo fraction in clustered and
unclustered low-mass stars are derived assuming the stars have
zero metallicity and that the halo density $\rho$ varies with
Galactocentric cylindrical coordinates ($R,z$) as
   \be
      \rho = \rho_0 \left( \frac{R_{\rm c}^2 + R_0^2}{R_{\rm c}^2 +
      R^2 + z^2} \right) \label{e1}
   \ee
where, in Paper~I, the local density $\rho_0 = 0.01~\sm$~pc$^{-3}$,
the Solar Galactocentric distance $R_0 = 8$~kpc, and the halo core
radius $R_{\rm c} = 5$~kpc.

The assumption of zero metallicity is maintained in the present
analysis since one expects the halo to be perhaps the oldest of the
Galactic components, and hence its constituents to have more or less
primordial metallicity. The expected absolute magnitude in various
photometric bands for such stars between the hydrogen-burning limit
mass ($0.092~\sm$) and $0.2~\sm$ has been calculated by Saumon et
al. (\cite{sau94}) and their results are employed here as in Paper~I.

The model dependency of the conclusions in Paper~I is
assessed by re-calculating the constraints for a number of different,
but plausible, halo models. For ease of comparison the models selected
are 5 of the reference halo models used by the MACHO collaboration
(Alcock et al. \cite{alc96}) in its analysis. (MACHO considers a total
of 8 Galactic models, though only 5 of the halo models have distinct
functional forms.) All halo models assume $R_0 = 8.5$~kpc and $R_{\rm
c} = 5$~kpc. The 5 models are denoted by MACHO as models A--D and S
(for `standard'), and this labelling is maintained here.

The standard model S has the same functional form as the halo
investigated in Paper~I (i.e. it is described by Eq.~\ref{e1}) but uses
the slightly larger IAU value for $R_0$ above and assumes a lower
local density $\rho_0 = 0.0079~\sm$~pc$^{-3}$. Models A--D are drawn
from the self-consistent family of power-law models (Evans
\cite{evan94}), having density profiles
   \begin{eqnarray}
      \rho & \!\! = \!\! & \frac{v_{\rm a}^2 R_{\rm c}^\beta}{4 \pi G q}
      \frac{R_{\rm c}^2(1 + 2q^2) + R^2(1 - \beta q^2) + z^2[2
      - q^{-2}(1+\beta)]}{(R_{\rm c}^2 + R^2 + z^2
      q^{-2})^{(\beta+4)/2}}, \nonumber \\ & & \label{e2}
   \end{eqnarray}
where $v_{\rm a}$ is the velocity normalisation, $q$ describes the
flattening of equipotentials, $\beta$ determines the power-law slope
of the density profile at large radii, and $\pi$ and $G$ have their
usual meanings. For a flat rotation curve at large radii $\beta = 0$,
where as for a rising curve $\beta < 0$ and for a falling one $\beta >
0$.

\setcounter{table}{0}
\begin{table}
\caption{Parameter values for the 5 MACHO reference halo models A--D
and S (Alcock et al. \cite{alc96}). $R_0 = 8.5$~kpc and $R_{\rm c} =
5$~kpc is assumed for all models. For models A--D the local density
$\rho_0$ is derived from the parameters in columns 2--4. The
local rotation speed $v_0$ is computed from the combined halo and disc
mass within $R_0$.}
\label{t1}
\begin{tabular}{cccccc}
\hline\noalign{\smallskip}
Model & $v_{\rm a}$/km~s$^{-1}$ & $q$ & $\beta$ &
$\rho_0/\sm$~pc$^{-3}$ & $v_0$/km~s$^{-1}$ \\
\noalign{\smallskip}\hline\noalign{\smallskip}
A & 200 & 1 & 0.0 & 0.0115 & 224 \\
B & 200 & 1 & -0.2 & 0.0145 & 233 \\
C & 180 & 1 & 0.2 & 0.0073 & 203 \\
D & 200 & 0.71 & 0.0 & 0.0190 & 224 \\
S & -- & -- & -- & 0.0079 & 192 \\
\noalign{\smallskip}\hline
\end{tabular}
\end{table}

The particular parameters for models A--D, along with those of model S
are listed in Table~\ref{t1}. Model A is the closest analogy to model
S within the power-law family of models, whilst model B has a rising
rotation curve at large radii, model C a falling rotation curve, and
model D a flattening equivalent to an E6 halo. When combined with the
MACHO canonical Galactic disc (Alcock et al. \cite{alc96}), the models
give values for the local Galactic rotation speed $v_0$ within 15\% of
the IAU standard value of 220~km~s$^{-1}$ and have rotation curves
that are consistent with observations.

\section{HST observations and halo fraction constraints}

Gould et al. (\cite{gou97}) have calculated the disc luminosity
function for M-dwarf stars using data from several HST WFC2
fields. These include 22 fields originally analysed by Gould et
al. (\cite{gou96}), along with the Hubble Deep Field, 28 overlapping
fields comprising the Groth Strip, and 2 other new fields: a total of
53 WFC2 fields. In Paper~I, 20 of the original 22 fields are analysed,
the other 2 fields being omitted due to statistical problems
introduced by their close proximity to some of the other fields
(namely that clusters appearing in these fields could also appear in
the other fields and thus be double counted). In this study these 20
fields are combined with the new fields analysed by Gould et
al. (\cite{gou97}), making the total number of fields 51.  The
nearest-neighbour separation between these fields is sufficiently
large that double counting is not expected to be a problem for
clusters of interest. (The overlapping Groth Strip fields are treated
as a single large field for the purpose of this study.)

The limiting and saturation $I$-band magnitudes for the fields are
listed in Table~1 of Gould et al. (\cite{gou97}). The Groth Strip is
treated as a single field with an angular coverage of 25.98 WFC2 fields
(this accounts for overlaps) and magnitude limits corresponding to the
modal values listed in Gould et al. (\cite{gou97}). As in Paper~I,
these limits are translated into star-mass dependent limiting
distances by converting the line-of-sight extinction values listed in
Burstein \& Heiles (\cite{bur84}) to $I$-band reddenings and using the
photometric predictions of Saumon et al. (\cite{sau94}) for
zero-metallicity low-mass stars. The predictions for the $V$ and $I$
bands are well fit by the colour--magnitude relation
   \be
      I = -11.45 \, (V-I)^2 + 40.7 \, (V-I) - 24.5 \label{e2.5}
   \ee
for $1.27 \le V-I \le 1.57$ (corresponding to $0.2 \ge m/\sm \ge 0.092$).  

The analyses for the unclustered and clustered scenarios proceed as in
Paper~I, except that the models listed in Table~\ref{t1} of this paper
now replace the model used there. The calculations for the cluster
scenario, which are described in detail in Paper~I, assume that the
surface-brightness profiles of the clusters follow the King
(\cite{king62}) surface-brightness law and take into account cluster
resolvability, as well as line-of-sight overlap.

\setcounter{table}{1}
\begin{table}
\caption{Constraints on unclustered zero-metallicity low-mass stars in
the Galactic halo arising from the detection of 145 candidate stars
within 51 HST WFC2 fields. The second and third columns give the
expected number of detectable stars $N_{\rm exp}$ for a full halo
($f_{\rm h} = 1$) for stars with masses of $0.2~\sm$ and $0.092~\sm$
(the hydrogen-burning limit mass), respectively. The last two columns give
the 95\% confidence upper limit on the maximum halo fraction $f_{\rm
max}$.}
\label{t2}
\begin{tabular}{ccccc}
\hline\noalign{\smallskip}
      & \multicolumn{2}{c}{$N_{\rm exp}$} 
      & \multicolumn{2}{c}{$f_{\rm max}$} \\
      \noalign{\smallskip}\cline{2-5}\noalign{\smallskip}
Model & $0.2~\sm$ & $0.092~\sm$ & $0.2~\sm$ & $0.092~\sm$ \\
\noalign{\smallskip}\hline\noalign{\smallskip}
A & 183\,000 & 24\,100 & $9 \times 10^{-4}$ & 0.007 \\
B & 248\,000 & 30\,800 & $6 \times 10^{-4}$ & 0.005 \\
C & 109\,000 & 15\,000 & 0.0015 & 0.011 \\
D & 162\,000 & 34\,300 & 0.0010 & 0.005 \\
S & 141\,000 & 17\,100 & 0.0012 & 0.010 \\
\noalign{\smallskip}\hline
\end{tabular}
\end{table}

Table~\ref{t2} lists the results for the unclustered scenario. Within
the 51 HST WFC2 fields analysed a total of 145 candidate stars with
$1.2 \le V-I \le 1.7$ are found, implying a 95\% confidence level (CL)
upper limit on the average number of 166 stars. This colour range
spans the $V-I$ colour predictions of Saumon et al. (\cite{sau94}) for
stars with masses in the interval $0.092-0.2~\sm$, where the lower
value corresponds to the hydrogen-burning limit. Comparison with the
expected number tabulated in Table~\ref{t2} clearly shows that, for all
models, even the lowest mass unclustered stars fall well short of
providing the halo dark matter density inferred by MACHO. The upper
limit on their fractional contribution $f_{\rm max}$ is shown for
0.2-$\sm$ and 0.092-$\sm$ stars. For the lowest mass stars $f_{\rm
max}$ ranges from 0.5\% for models B and D to 1.1\% for the lighter
halo model C.

One interesting feature of Table~\ref{t2} is that for the flattened
halo model D the expected number counts are enhanced for 0.092-$\sm$
stars relative to the predictions for the spherically symmetric
models, producing the highest predicted number-count for these
stars. This contrasts with the results for the brighter 0.2-$\sm$
stars, with the heavy halo model B producing the highest number-count
prediction. The enhancement for 0.092-$\sm$ stars in model D arises
because the flattening preferentially increases the stellar surface
density near the Galactic plane, and this is reflected in the counts of
0.092-$\sm$ stars which can be at most only a few kpc from the plane
if they are to be detected.

\begin{figure*}
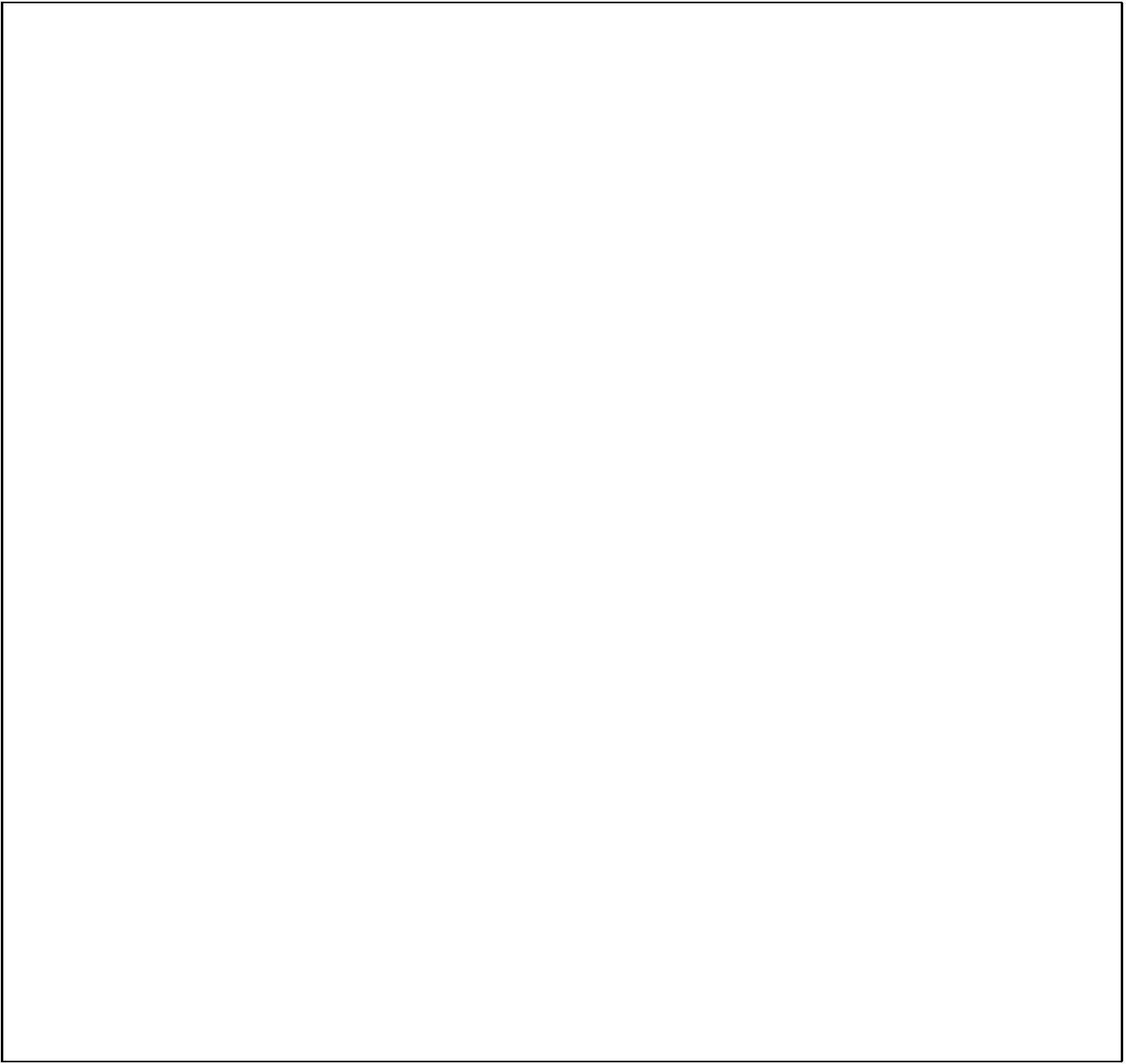

\picplace{17.0cm}
\caption[]{Comparison of constraints on the halo fraction $f_{\rm h}$
from HST limits, MACHO observations and dynamical constraints for the
5 halo reference models (A--D, S), assuming halo stars have a mass of
$0.092~\sm$ and all reside in clusters with mass $M$ and radius
$R$. The lower plateau to the left of each plot corresponds to the
95\%~CL upper limit $f_{\rm max}$ for the unclustered scenario
inferred from HST counts (see Table~\ref{t2}). The upper plateau
on the right corresponds to the 95\%~CL lower limit halo fraction
$f_{\rm M,low}$ inferred by MACHO 1st- and 2nd-year observations
(Alcock et al. \cite{alc97}), with the central value for the MACHO
halo fraction $f_{\rm M}$ indicated by the skirting surrounding the
plots (see also Table~\ref{t3}). The curved surface joining the lower and
upper flat regions corresponds to the 95\%~CL upper limit on the halo
fraction in clusters from HST counts. Also projected onto the plane
$f_{\rm h} = f_{\rm M,low}$ are the cluster dynamical constraints
(dashed lines) for the local Solar neighbourhood. The intersection
between these constraints and the MACHO lower-limit plateau indicates
cluster parameters compatible with HST, MACHO and dynamical
constraints.}
\label{f1}
\end{figure*}

The constraints on the halo fraction $f_{\rm h}$ for the clustered
scenario as a function of cluster mass $M$ and radius $R$ are shown in
Fig.~\ref{f1} for the 5 models (A--D, S) assuming all stars reside in
clusters and have the hydrogen-burning limit mass of $0.092~\sm$. Each
plot is characterised by a lower plateau to the left, an upper plateau
to the right and a curved rising surface joining the two. This curved
surface between the two flat regions represents the 95\%~CL upper
limit halo fraction in clusters inferred from the presence of only 145
candidate stars within the 51 HST WFC2 fields. The constraints are
actually calculated on the basis of no stars being present within these
fields, since for clusters there is little difference in the
constraints assuming no stars are found or assuming a few hundred
stars are found. The reason for this, as discussed in Paper~I, is that
the clusters considered here contain between 1000 and $10^7$ members
each, so the presence of just one cluster within any of the HST fields
would typically result in thousands if not millions of candidates
being detected.

The lower plateau shows the 95\%~CL upper limit halo fraction for the
{\em unclustered \/} scenario (corresponding to the $f_{\rm max}$
values listed in Table~\ref{t2}). Clusters with masses and radii within
this region have internal densities which are lower than that of the
halo background average and are thus unphysical, since they represent
local under-densities rather than over-densities. Clearly constraints on
clusters cannot be stronger than constraints on a smooth stellar
distribution. The intersection of the lower plateau with the curved
rising surface therefore denotes the boundary between unphysical
and physical cluster parameters.

\setcounter{figure}{0}
\begin{figure}
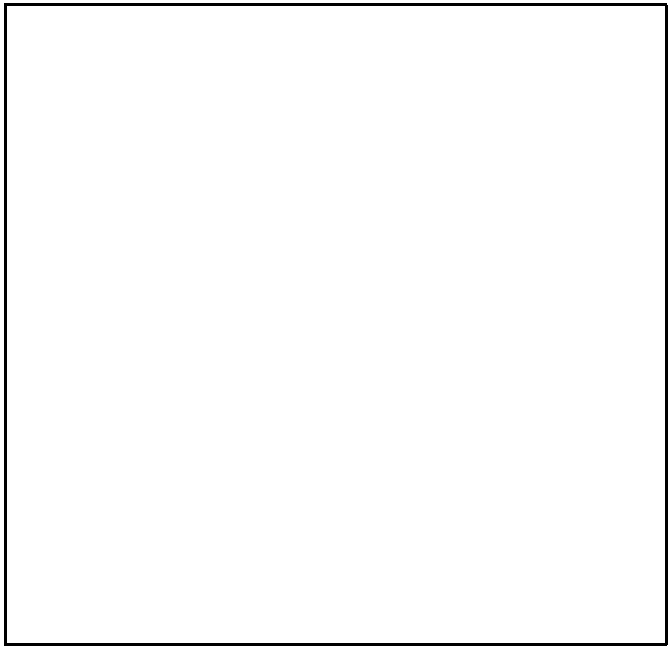

\picplace{8.5cm}
\caption[]{{\em continued.}}
\end{figure}

The upper plateau to the right represents the 95\%~CL {\em lower
limit\/} on the halo fraction $f_{\rm M,low}$ inferred from MACHO 1st-
and 2nd-year microlensing results (Alcock et al. \cite{alc97}). It is
calculated by taking the 95\%~CL lower limit on the measured
microlensing optical depth for all 8 MACHO events ($\tau > 1.47 \times
10^{-7}$), subtracting the optical depth contribution expected from
non-halo components [corresponding to $\tau_{\rm non-halo} \simeq 5
\times 10^{-8}$ (Alcock et al. \cite{alc96})], and normalising to the
optical depth prediction $\tau_{\rm exp}$ for a full halo ($f_{\rm h}
= 1$) for each model. The top of the skirting surrounding each plot is
normalised to the {\em central\/} MACHO value for the halo fraction
$f_{\rm M}$ for comparison, and is calculated in a similar manner to
the lower limit ($f_{\rm M}$ and $f_{\rm M,low}$, together with
$\tau_{\rm exp}$, are tabulated in Table~\ref{t3} for each
model). Since this plateau lies below the extrapolation of the HST
cluster-fraction constraint [which rises asymptotically over this
region -- see Fig.~2 of Paper~I], it is consistent with both MACHO and
HST limits.

The dashed lines in the plots of Fig.~\ref{f1} represent the dynamical
constraints derived for the local Solar neighbourhood. In fact some of
the HST fields are somewhat closer in to the Galactic centre, where
the dynamical constraints are stronger, but most are further away so
the limits shown are stronger than applicable for most of the HST
fields. The functional form for the constraints are detailed in
Paper~I and are dependent upon Galactic as well as cluster parameters
[consult Lacey \& Ostriker (\cite{lac85}); Carr \& Lacey (\cite{carr87});
Moore (\cite{moo93}); Moore \& Silk (\cite{moo95}); Carr \& Sakellariadou
(\cite{carr97}) for derivations, and see Carr (\cite{carr94}) for a
detailed review of dynamical constraints]. Their variation from plot
to plot is due to model variations in the local density and rotation
speed (see Table~\ref{t1}). The dynamical constraints are projected
onto the plane $f_{\rm h} = f_{\rm M,low}$ for direct comparison with
the MACHO lower limits. The intersection of the MACHO lower-limit
plateau with the dynamical limits therefore represents cluster
parameters compatible with MACHO, dynamical limits, and the
constraints from the 51 HST fields.

For each model it is evident that the region compatible with all
limits spans a significant range of masses and radii. For models C and
D the maximum permitted cluster mass is around $5 \times 10^4~\sm$,
whilst for models A, B and S one can have cluster masses in excess of
$10^6~\sm$. Interestingly, whilst in the unclustered scenario the
heavy halo model~B is the most strongly constrained in terms of
allowed halo fraction $f_{\rm max}$, it nonetheless allows a
relatively wide range of viable cluster masses in the clustered
scenario. Conversely, the permitted cluster mass range for the light
halo model~C is more restricted. 

This apparent paradox is due to the fact that the HST,
dynamical and microlensing observations limit the halo density
normalisation at different positions in the halo, so their
intersection is sensitive to the halo density profile. In particular,
the HST and dynamical limits essentially apply to the local Solar
neighbourhood position ($R_0 = 8$~kpc) for clusters comprising
relatively dim hydrogen-burning limit stars, where as the microlensing
observations towards the LMC constrain the density of lenses at
somewhat larger distances (primarily between 10 and 30~kpc from the
Galactic centre, where the product of lens number density and lensing
cross-section is largest). Hence, for a given microlensing constraint
on the mass density of lenses at 10 to 30~kpc, the local dynamical and
number-count constraints are weaker for haloes with rising rotation
curves (such as model B) than for models with falling rotation curves
(such as model C).

The relatively large range in allowed cluster masses and radii for
model~S is in apparent contrast to the results of Paper~I, in which
the surviving parameter space is shown to be much smaller for the very
similar model adopted there. There are two reasons for this apparent
discrepancy: (1) in Fig.~\ref{f1} of this paper it is assumed that the
clusters comprise hydrogen-burning limit stars, where as in Fig.~3 of
Paper~I the constraints are shown for the brighter 0.2-$\sm$ stars;
(2) in this study consistency is being demanded only with the {\em
lower\/} limit MACHO halo fraction $f_{\rm M,low}$, rather than with
the central value $f_{\rm M}$ as in Paper~I. This latter difference is
particularly important because it enlarges both the sizes of the
dynamically-permitted region and the MACHO plateau, and hence enlarges
their intersection.  Since these differences serve to maximise the
size of the surviving region, the constraints shown in this paper
should be taken as {\em firm\/} limits on allowed cluster parameters.

\section{Constraints on cluster membership}

Figure~\ref{f1} assumes that all stars reside in clusters at the
present day, an unrealistic assumption since one expects some fraction of
the clusters to have evaporated away over time. As in Paper~I one can
place limits on the fraction of stars $f_{\rm c}$ which must remain in
clusters by using the strong limits $f_{\rm max}$ on the
unclustered scenario (listed in Table~\ref{t2}). Assuming the lower limit
on the cluster halo fraction to be given by the lower limit inferred
by MACHO, $f_{\rm M,low}$, the present-day halo fraction in stars
which have evaporated away from clusters is $f_{h,*} > (1-f_{\rm
c})f_{\rm M,low}$. Since HST observations demand $f_{h,*} \leq f_{\rm
max}$ one has
   \be
      f_{\rm c} > 1 - (f_{\rm max}/f_{\rm M,low}). \label{e3}
   \ee
The resulting values for $f_{\rm c}$ for 0.2-$\sm$ and 0.092-$\sm$
stars are given in Table~\ref{t3}.

\setcounter{table}{2}
\begin{table}
\caption{Microlensing halo fractions and minimum clustering fractions
for the reference halo models. Column~2 gives the expected optical
depth for a full halo as calculated by Alcock et
al. (\cite{alc96}). Column~3 gives the central value for the halo
fraction using the 1st+2nd year optical depth estimate of $2.94 \times
10^{-7}$ measured by Alcock et al. (\cite{alc97}), and subtracting
from it an optical depth of $5 \times 10^{-8}$ expected from non-halo
populations. The 4th column gives the 95\%~CL lower limit on the halo
fraction using the lower limit for the measured optical depth of $1.47
\times 10^{-7}$. The last two columns give the lower limit on the
present-day clustering fraction using column~4, Eq.~\ref{e3} and
Table~\ref{t2}.}
\label{t3}
\begin{tabular}{cccccc}
\hline\noalign{\smallskip}
  & & & & \multicolumn{2}{c}{$f_{\rm c}$} \\
  \noalign{\smallskip}\cline{5-6}\noalign{\smallskip}
Model & $\tau_{\rm exp}/10^{-7}$ & $f_{\rm M}$ & $f_{\rm M,low}$ &
$0.2~\sm$ & $0.092~\sm$ \\
\noalign{\smallskip}\hline\noalign{\smallskip}
A & 5.6 & 0.43 & 0.17 & 0.995 & 0.96 \\
B & 8.1 & 0.30 & 0.12 & 0.995 & 0.96 \\
C & 3.0 & 0.81 & 0.32 & 0.995 & 0.97 \\
D & 6.0 & 0.41 & 0.16 & 0.994 & 0.97 \\
S & 4.7 & 0.52 & 0.21 & 0.994 & 0.95 \\
\noalign{\smallskip}\hline
\end{tabular}
\end{table}

From Table~\ref{t3} it is clear that all models require a very high
fraction of all stars to reside in clusters at present. Even for
hydrogen-burning limit stars the required clustering fraction must be
at least 95\% at present. Capriotti \& Hawley (\cite{cap96}) have
undertaken a detailed analysis of cluster mass loss within
an isothermal halo potential for a range of cluster masses, density
profiles and Galactocentric distances. Their analysis takes account of
evaporation, disruption and tidal processes. They find that clusters
with masses between $10^5 - 10^7~\sm$ generally survive largely intact
to the present day but that less massive clusters survive only if they
have high central density concentrations, nearly circular orbits and
reside at large distances from the Galactic centre. At the Solar position
Capriotti \& Hawley find that clusters with a half-mass to tidal radii
ratio of 0.3 (comparable to the value for the clusters analysed here
and in Paper~I) survive more than 95\% intact only if they have masses
exceeding $10^6~\sm$. 

However, there are a number of reasons why these limits may be
stronger than applicable to the low-mass star cluster
scenario. Firstly, Capriotti \& Hawley assume that the clusters
comprise $m = 0.8~\sm$ stars (i.e. between 4 and 9 times more massive
than the stars considered in the present study). The evaporation
timescale scales approximately as $m^{-1}$ for a fixed cluster mass,
so for clusters comprising lower mass stars the evaporation timescale
is correspondingly longer. Secondly, the local halo density assumed by
Capriotti \& Hawley of $0.0138~\sm$~pc$^{-3}$ at a Galactocentric
distance of 8.5~kpc is on the higher end of the values for the halo
models analysed in this paper, and is considerably larger than the
allowed MACHO lower limit, $f_{\rm M,low} \rho_0$, on the local
density in lenses (by a factor of between 5 and 10 after normalising
to a distance $R_0 = 8$~kpc). Hence disruption due to close encounters
with other clusters is substantially less in the halo models
investigated here than for the model analysed by Capriotti \& Hawley.

Lastly, a study by Oh \& Lin (\cite{oh92}) has shown that the cluster
escape rates may be substantially smaller than commonly assumed due to
angular momentum transfer arising from the action of the Galactic
tidal torque on cluster stars with highly eccentric orbits (which in
the absence of the torque would constitute the bulk of the
escapees). The rates calculated by Oh \& Lin for isotropic cluster
models are broadly consistent with the values used by Capriotti \&
Hawley (\cite{cap96}) and other authors, but for the case of
anisotropic stellar orbits the escape rates can be 1--2 orders of
magnitude smaller, again implying correspondingly longer evaporation
timescales. It therefore appears that, under certain conditions, one
may be able to reconcile the high cluster fraction requirements derived
in the present study with the findings of cluster dynamical studies,
at least for clusters comprising stars close to the hydrogen-burning
limit.

In any case, the validity of the figures in Table~\ref{t3}
depend upon just how smoothly distributed are the stars which have
evaporated from clusters. If they still have not completely
homogenised today, instead maintaining a somewhat lumpy distribution
(reflecting their cluster origin), then the limits on $f_{\rm c}$ are
too strong.

For example, a cluster with a mass $3 \times 10^4~\sm$ and radius 3~pc
represents an over-density of about $3 \times 10^4$ over the
background average at the Solar neighbourhood [i.e $\delta \rho/\rho
\equiv (\rho - \overline{\rho})/\overline{\rho} = 3 \times
10^4$]. However, an under-density in the unclustered (or more
precisely `post-clustered') stellar population of just a factor 10
($\delta \rho/\rho = -0.9$) over volumes larger than $3\times
10^5$~pc$^3$, which is roughly the volume probed by 50 HST fields for
hydrogen-burning limit stars (and is of order 10 times smaller than
the halo volume per cluster), is all that is required to weaken the
constraints on $f_{\rm max}$ by a factor 10. This would result in a
much more comfortable lower limit on $f_{\rm c}$ of just 0.5 for
0.092-$\sm$ stars, rather than 0.95. If the under-density is a factor
5 lower than the background ($\delta \rho/\rho = -0.8$) one requires
$f_{\rm c} > 0.75$ for the lowest mass stars and for an under-density
factor of 2 ($\delta \rho/\rho = -0.5$) $f_{\rm c}$ must exceed 0.9.

In order to rule out the cluster scenario definitively (say with 95\%
confidence) one needs a survey that is both sufficiently wide and deep
that it might be expected to contain at least 3 clusters on average,
regardless of their mass and radius (though their mass and radius must
be dynamically permitted). From Fig.~\ref{f1} it appears that the most
difficult dynamically-allowed clusters for HST to exclude are those
with a mass of around $3 \times 10^4~\sm$. If the halo fraction in
low-mass stars is around 40\%, typical of the preferred value for the
MACHO results, then the local number density of such clusters is around
130~kpc$^{-3}$ (adopting a local halo density of $0.01~\sm$~pc$^{-3}$;
in reality of course the average density within the fields is
dependent upon the halo model and the field locations). If the
clusters comprise hydrogen-burning limit zero-metallicity stars ($V-I
= 1.57$) then a HST-type survey will be sensitive to them out to about
3.6~kpc in the $I$ band [using the colour-magnitude relation of
Eq.~(\ref{e2.5}), and assuming a limiting $I$-band sensitivity of
24~mag], so in order to expect to detect at least 3 such clusters, the
survey must cover a solid angle of at least 4.5~deg$^2$, or the
equivalent of 3\,700 HST fields!  Therefore, only if HST fails to
detect any clusters from 3\,700 fields would dynamically-allowed
clusters be ruled out with 95\% confidence from explaining all of the
observed microlensing events. For comparison, an all-sky $K$-band
survey over Galactic latitudes $|b| > 10 \degr$ requires a
limiting magnitude of about 17.5 in order to produce similar
constraints. This should be compared to the expected $K$-band limit of
about 14 for the ground-based DENIS and 2MASS surveys.

An easier alternative is to instead obtain several fields as close to
the Galactic centre as is feasible, where the dynamical constraints
are much stronger than for the Solar neighbourhood position. For
low-mass stars, this necessitates a telescope such as HST with the
capability for obtaining very deep fields, since shallow surveys with
wide angular coverage essentially only probe the local Solar
neighbourhood.

\section{Conclusion}

Kerins (\cite{ker97}), referred to as Paper~I, has suggested that low
mass stars could provide the substantial dark matter fraction
indicated by the combined 1st- and 2nd-year MACHO gravitational
microlensing results. Whilst observations from Hubble Space Telescope
(HST) and other instruments have been interpreted as excluding such
stars from having a significant halo density, Paper~I shows that their
density could in fact be substantial if they are grouped into
globular-cluster configurations. The motivation for such clusters
comes from the baryonic dark matter formation scenarios which are
discussed in Paper~I. Paper~I calculates the constraints on such
clusters (assuming they comprise low-mass stars of primordial
metallicity) which arise from MACHO microlensing results, dynamical
constraints on massive halo objects, and observations from 20 HST
fields obtained by Gould et al. (\cite{gou96}). However, the results
of Paper~I apply only to the spherically-symmetric cored isothermal
halo model investigated there.

In the present study, the number of HST fields utilised has been
increased to 51, and now incorporates the Hubble Deep Field and Groth
Strip fields (Gould et al. \cite{gou97}). The model dependency of the
results in Paper~I has been tested by adopting 5 of the reference halo
models employed in the MACHO collaboration's analysis of its
microlensing results. One of the models is similar to the halo
investigated in Paper~I whilst the other 4 are drawn from a
self-consistent family of power-law halo models and comprise
spherically-symmetric haloes with a rising rotation curve, a falling
rotation curve and a flat curve, as well as a flattened (E6) halo
model.

The 51 HST fields contain just 145 candidates with $V-I$ colours
between 1.2 and 1.7 (spanning the colour range predicted for
zero-metallicity stars with masses between the hydrogen-burning limit
and $0.2~\sm$) against the tens or hundreds of thousands predicted for
the halo models. From this one concludes that the halo fraction in
unclustered low-mass stars is at most $0.5-1.1\%$ with 95\%
confidence, depending on the halo model, and in all cases falls well
short of providing even the lower-limit halo fraction inferred by
MACHO.

However, in the cluster scenario there exists a wide range of cluster
masses and radii which can allow a halo fraction consistent with the
lower limit derived from MACHO microlensing results whilst remaining
compatible with dynamical limits and HST observations. Consistency
with the preferred microlensing halo fraction, rather than the lower
limit, requires fine tuning of the cluster parameters (as found in
Paper~I), but is possible for all models investigated.

The one potentially serious problem for the cluster scenario is that
the strong constraints on unclustered stars imply that an overwhelming
fraction of all stars, at least 95\%, must still reside in clusters at
the present day. This is higher than expected from generic cluster
evaporation considerations for much of the permitted cluster mass
range, though it may still be consistent with clusters comprising
stars with anisotropic orbits. In any case, these limits assume that
stars which have already evaporated from clusters now form a perfectly
smooth distribution which traces the halo density profile. If instead
these stars still have a lumpy distribution, reflecting the fact that
they previously resided in clusters, then the cluster fraction limits
are too strong.

Probably the only way to definitively exclude or confirm the cluster
scenario is to obtain several deep fields as close to the Galactic
centre as is practical, where the strong dynamical constraints
severely restrict the range of feasible cluster parameters.

\begin{acknowledgements}
I am grateful to Andy Gould, John Bahcall and Chris Flynn for their
permission to use the HST data in advance of publication. This
research is supported by a EU Marie Curie TMR Fellowship.
\end{acknowledgements}

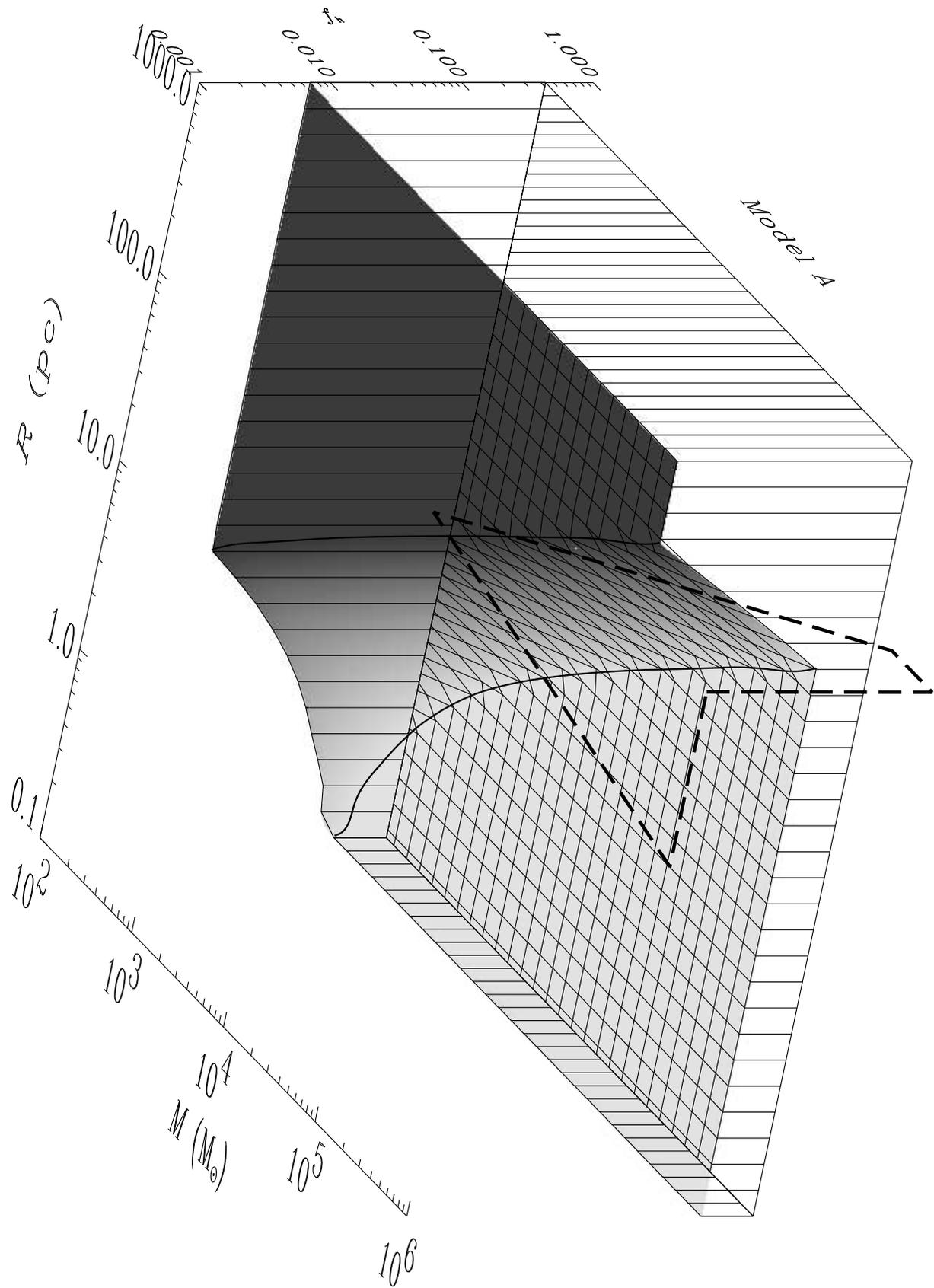
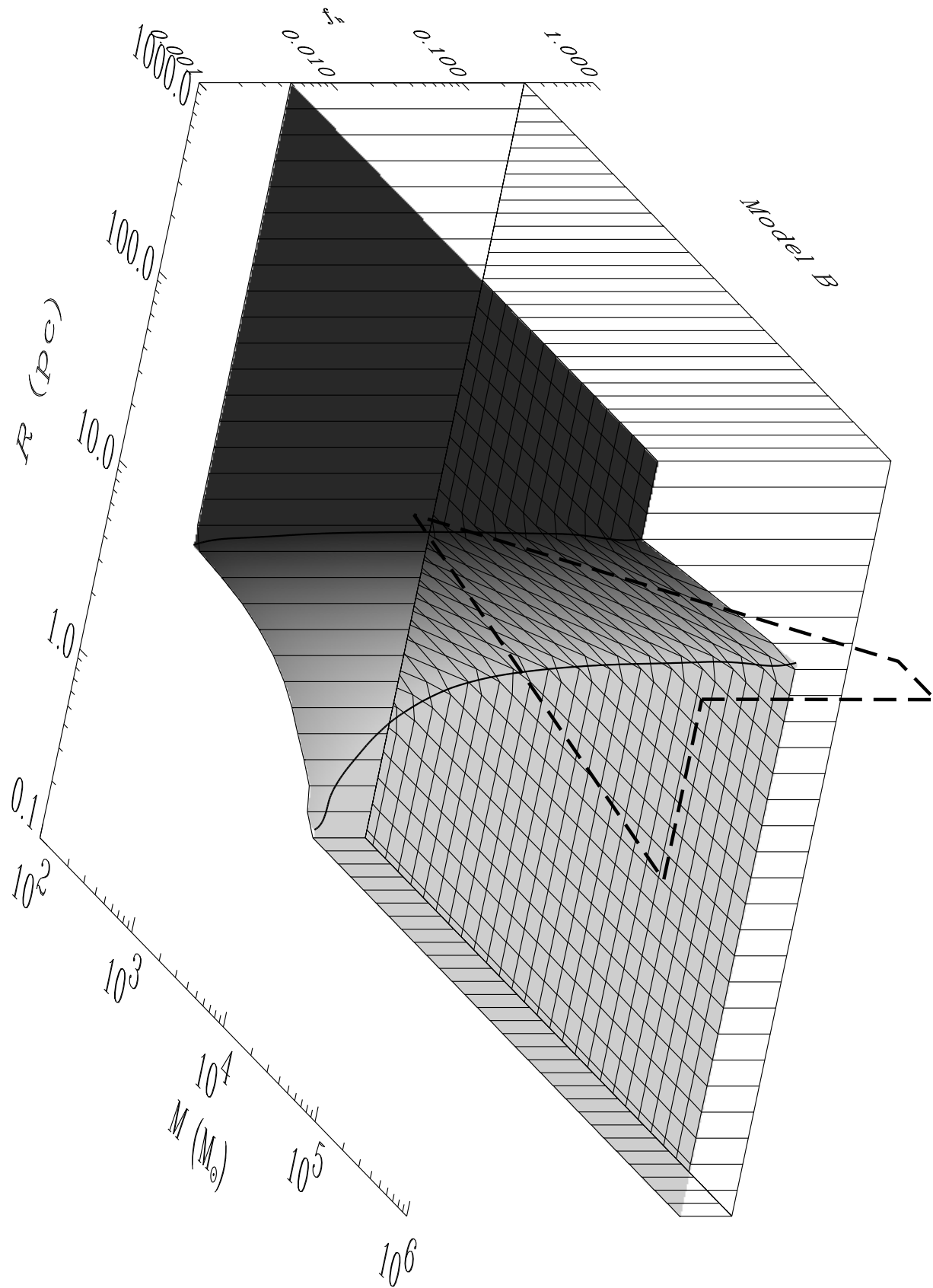
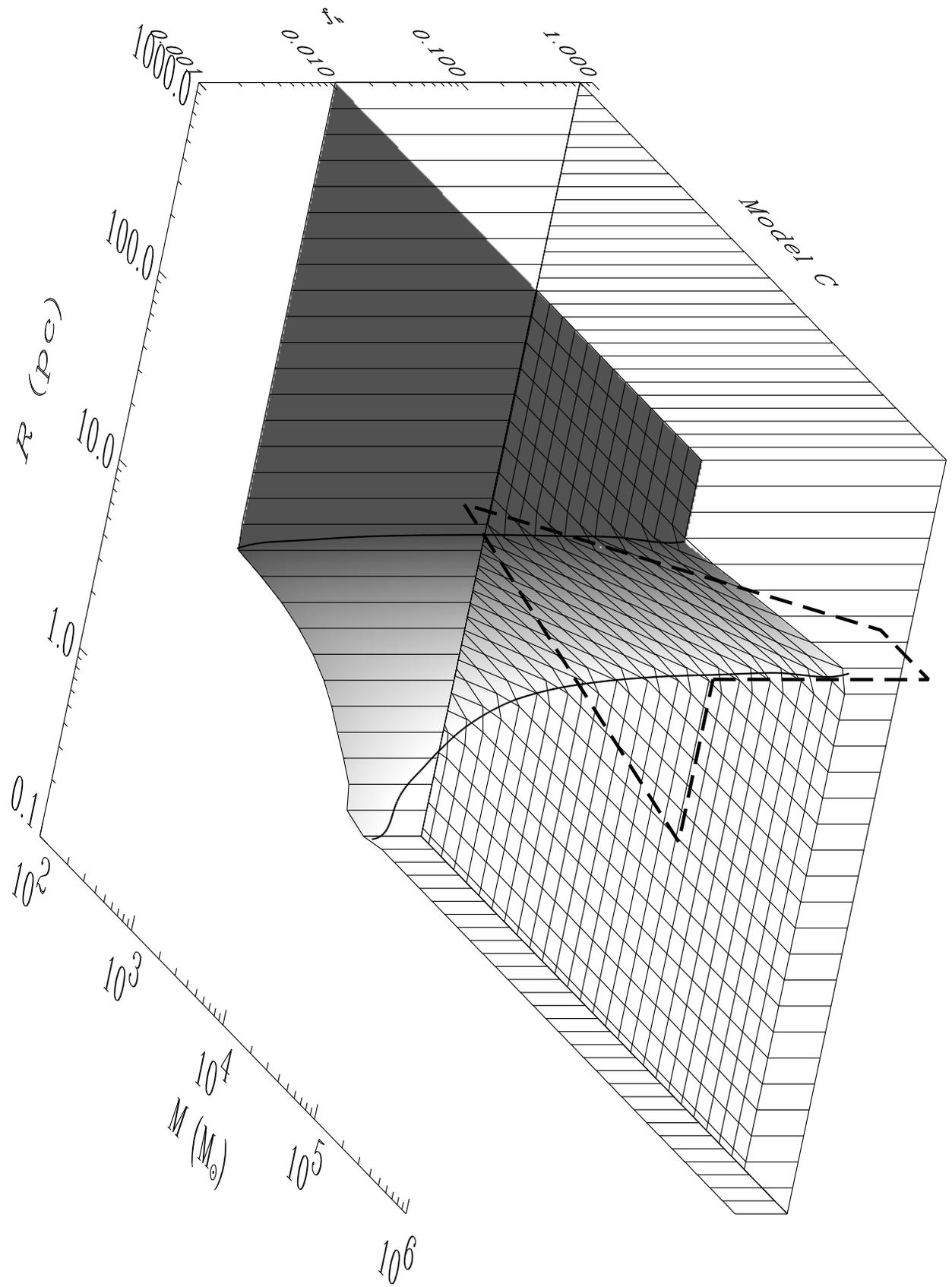
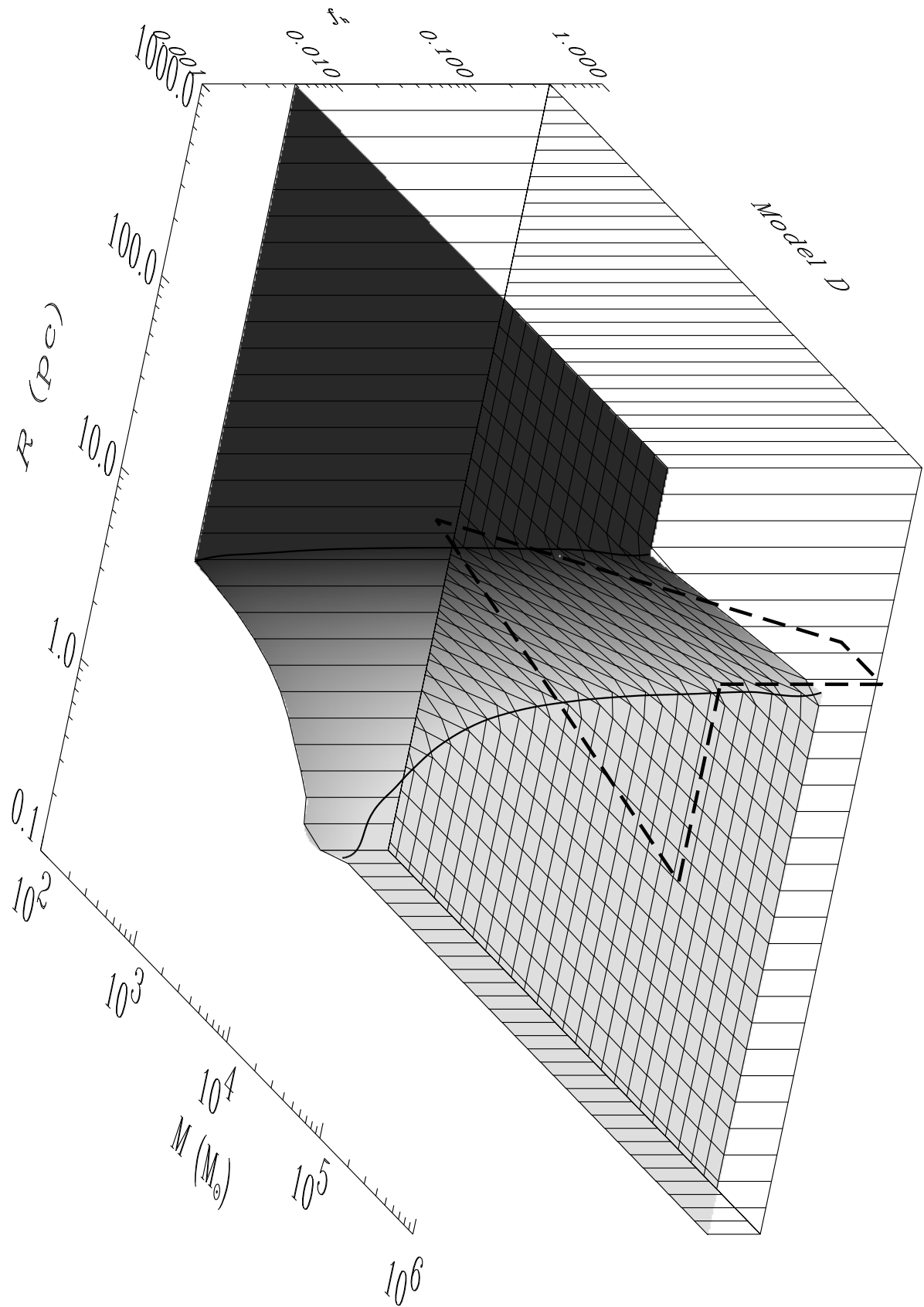
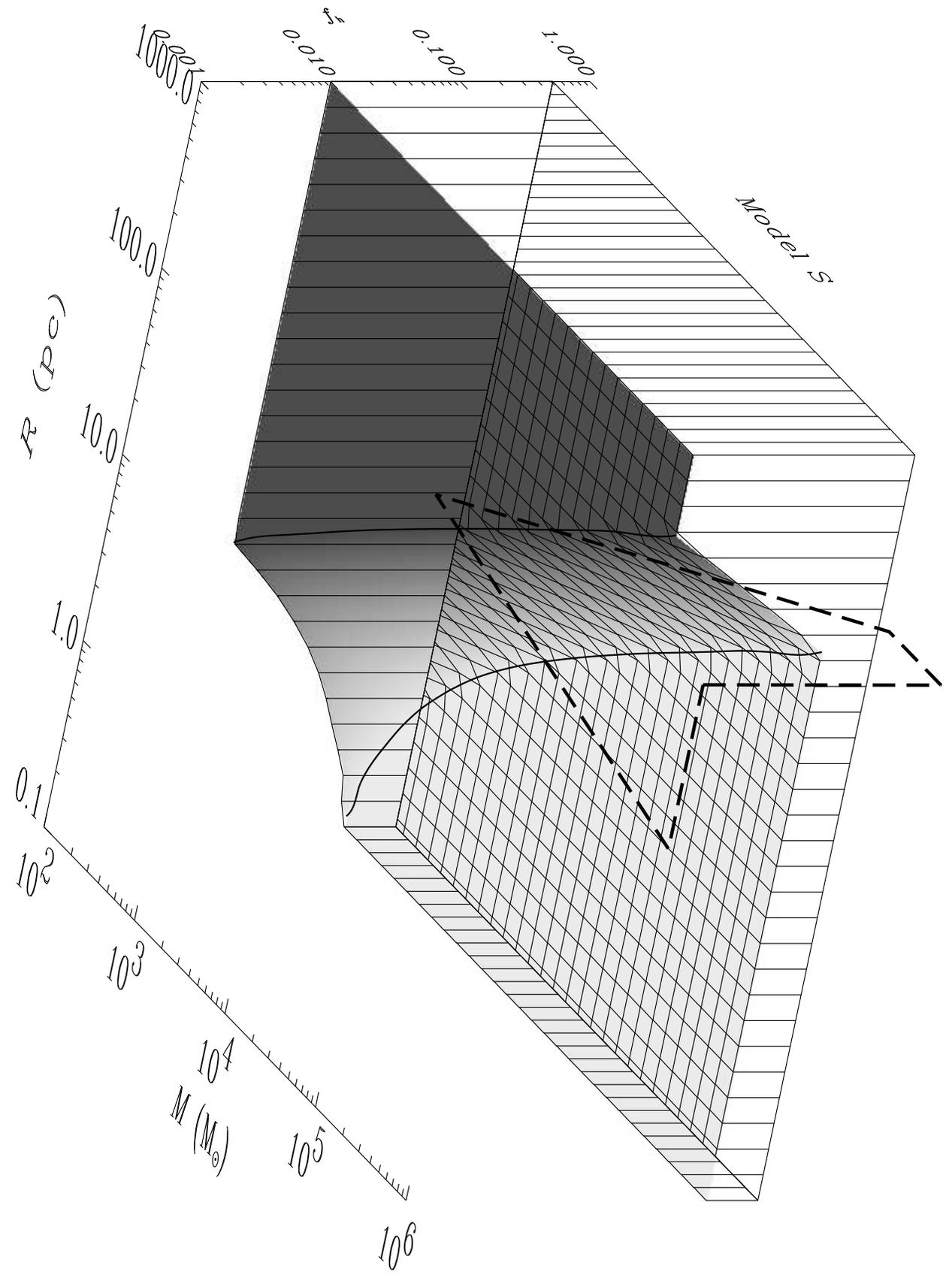
\end{document}